\documentstyle[aps,multicol,psfig,amsfonts]{revtex}

\begin{document}   
\newcommand{\be}{\begin{equation}}
\newcommand{\ee}{\end{equation}}
\newcommand{\br}{\bbox{r}}
\newcommand{\brp}{\bbox{r}^\prime}
\newcommand{\rd}{{\rm d}}
\newcommand{\re}{{\rm e}}
\newcommand{\ri}{{\rm i}}
\newcommand{\au}{\mbox{\,[a.u.]}}
\newcommand{\wc}{\omega_{\rm c}}
\renewcommand{\wp}{\omega_{\rm p}}
\newcommand{\rs}{r_{\rm s}}
\newcommand{\ks}{k_{\rm s}}
\newcommand{\qs}{q_{\rm s}}
\bibliographystyle{prsty}
\widetext  
\title{  
Collective versus single-particle effects
in the optical spectra of
finite electronic quantum systems}
\author{M. Santer and B. Mehlig}
\address{Theoretical Quantum Dynamics, University
         of Freiburg, Hermann-Herder-Str. 3, Freiburg}  
\date{\today}
\maketitle{ } 
\begin{abstract}  
We study optical spectra of finite electronic quantum
systems at frequencies smaller than the
plasma frequency
using a quasi-classical approach. 
This approach
includes collective effects and enables
us to analyse how the nature of the (single-particle) electron dynamics
influences the optical spectra in finite electronic
quantum systems.  We derive an analytical expression for the low-frequency 
absorption coefficient of electro-magnetic radiation
in a finite quantum system with ballistic
electron dynamics and specular reflection 
at the boundaries: a two-dimensional electron gas confined to
a strip of width $a$ (the approach can be applied to systems of any shape and 
electron dynamics -- diffusive or ballistic, regular or irregular motion). 
By comparing with results of numerical computations using the random-phase
approximation we show that our analytical
approach provides a qualitative and
quantitative understanding of the optical spectrum.
\end{abstract}   
\begin{multicols}{2}
Optical spectra of finite metallic systems
have been intensively investigated for
almost a century. Early approaches
such as Mie's \cite{mie08} are of classical nature.
In \cite{mie08} the absorption of electro-magnetic radiation
by  conducting spheres is determined.
It is shown that the absorption spectrum
exhibits a resonance at $\omega_{\rm p}/\sqrt{3}$
(where $\omega_{\rm p}$ is the bulk plasma frequency)
due to collective motion of the charge carriers.

In the last decades there has been a substantial
amount of work on the nature of such collective resonances in
metallic clusters \cite{eka99}, nuclei \cite{ber94}, thin films \cite{fer58,kan61,new70},
small metal particles \cite{carr}, and dimensionally
reduced quantum systems \cite{all83,fet86,heit92} investigating, 
in particular, quantum-mechanical
effects. In most of these cases,
the electron dynamics is {\em ballistic} (the mean free path $\ell$
is larger than the  system size $a$). 
 The majority of theoretical papers
on the Mie resonance of finite metallic
systems use the so-called random phase approximation (RPA),
a self-consistent, quantum-mechanical approach
incorporating collective effects.
The nature of the Mie resonance in finite
electronic quantum systems is well understood, 
both qualitatively and quantitatively
(for a recent review see for instance \cite{eka99}).

The emergence of the field of Mesoscopic Physics
has fueled an increased
interest in the electronic and optical properties
of finite, disordered quantum systems (with {\em diffusive} electron
dynamics, $\ell \ll a$) in external fields.
In this context, attention has largely focussed on
\mbox{(quasi-)static} properties, $\omega \simeq \Delta/\hbar$,
where $\Delta$ is the mean level spacing of the system
in question (which is generally smaller than $\omega_{\rm p}$
by many orders of magnitude). Mesoscopic fluctuations
of the static polarisability and the capacitance,
for instance, were characterised in \cite{efe96},
and the electron density itself in \cite{agam}.
In all of these cases,  collective effects
must be taken into account in order to adequately
deal with the screening of the external field.
In the static limit, a Thomas-Fermi (TF) {\em ansatz} is appropriate
\cite{efe96,agam}.

Much less is known about  optical spectra
in the low-frequency region $\Delta/\hbar < \omega \ll \omega_{\rm p}$,
despite the fact that this regime is of particular interest:
one expects that the spectra strongly depend
on the nature of the (single-particle) electron
dynamics: In {\em ballistic} systems, for example,
it was argued \cite{aus93} that optical spectra should
exhibit {\em resonances near multiples
of $\wc = \pi v_{\rm F}/a$}
(see also \cite{meh97e}).
Usually $\Delta/\hbar \ll \wc \ll \wp$ .
According to \cite{aus93}, these resonances
should overlap to give a {\em linear
frequency dependence} of the absorption
coefficient for $\omega \gg \wc$ in two-dimensional systems 
(classical and local electro-magnetic theories predict a quadratic dependence
independent of dimension). These
are striking and unexpected results. They were, 
however, obtained within a TF approximation
which is valid in the {\em static} limit.
It must be examined to which extent {\em dynamic} screening effects may modify 
the results \cite{note3}.

Furthermore, in systems with {\em ballistic chaotic} dynamics,
the optical spectra should, at least within a single-particle picture,
reflect universal energy-level correlations found in classically 
chaotic quantum systems\cite{boh}. This issue
was addressed in the 
seminal paper by Gorkov and Eliashberg \cite{gor65},
investigating the polarisability of metallic
particles with disordered walls.
However, the treatment neglects 
screening effects all together \cite{ric73,siv87,bla96}.
How does {\em dynamical} screening modify this picture?

Numerical calculations (based, for instance on the RPA)
are ill-suited to answer these questions: They provide little 
 qualitative insight 
in the low-frequency region and, more importantly, it is necessary
to consider small systems or to make use of symmetries in order to 
make the numerical computations feasible. 
Disordered and (asymmetric) chaotic systems are 
very difficult to deal with.
In order to understand how the classical single-particle
dynamics is reflected in optical spectra
of finite quantum systems, it is thus greatly desirable to
have an analytical theory incorporating collective effects.

The main result of this paper
is an analytical expression for
the absorption properties
for a two-dimensional electron gas confined to a strip 
of width $a$ [eqs.~(\ref{eq:absresult}) and (\ref{eq:result})].
This example exhibits all features of more complicated geometries.
However, in the case discussed here, numerical RPA calculations are
feasible and allow us to discuss the accuracy of
the analytical approach.

In the following it is assumed
that $\Delta/\hbar < \omega \ll \wp,E_{\rm F}/\hbar$
where $E_{\rm F}$ is the Fermi energy. It is furthermore
assumed that $\lambda,\lambda_{\rm s} \gg a  \gg\lambda_{\rm F}$
where $\lambda$ is the wave length of the external
radiation, $\lambda_{\rm s}$ is the skin depth
and $\lambda_{\rm F}$ is the Fermi wave-length.

We consider a closed metallic quantum system
placed in an external electric field
$\bbox{E}_{\rm ext}$ [with a time dependence
$\propto \exp(\ri\omega t)$]. If the  wave-length is
much larger than the system size $a$, the external field is 
approximately constant 
throughout the system and
(neglecting retardation effects) can be
written as the gradient of an electric potential
$\varphi_{\rm ext}(\br) = E_0 x \hat{\bf e}_x$
[$\br = (x,y,z)$ is a three-dimensional coordinate vector].
Within the RPA,
the electronic response of the system to $\bbox{E}_{\rm ext}$
is calculated by solving a set of self-consistent
equations for the effective electrical potential
$\varphi(\br) = \varphi_{\rm ext}(\br) + \delta\varphi(\br)$
[$\delta\varphi(\br)$ is the potential
due to the induced charge density $\delta\varrho(\br)$] 
\begin{eqnarray}
\label{eq:rpa}
\delta\varphi(\br) &=& \int\rd\br^\prime G(\br,\brp)\delta\varrho(\brp)\,,\\
\delta\varrho(\br) &=& -\int\rd\br^\prime \Pi_0(\br,\br^\prime;\omega)\,
\varphi(\br)     
\nonumber
\end{eqnarray}
with the boundary condition that $\delta\varphi(\br)$ vanishes
as $|\br| \rightarrow \infty$. $G(\br,\brp)$ is the
Green function of the Laplace equation
$\Delta G(\br,\brp) = -\epsilon_0^{-1} \delta(\br-\brp)$,
$\epsilon_0$ is the dielectric constant.
$\Pi_0(\br,\br^\prime;\omega)$ 
is the non-local polarisability
\begin{eqnarray}
\nonumber
\Pi_0(\br,\br^\prime;\omega)
&=& -2 e^2\sum_{\alpha,\beta} 
\frac{f(\varepsilon_\alpha)-f(\varepsilon_\beta)}{%
\varepsilon_\alpha\!-\!\varepsilon_\beta\!-\!\hbar\omega\!+\!\ri\gamma}
\\
&&\hspace*{1cm}\times\psi_\alpha^\ast(\brp)\psi_\beta(\brp)
\psi_\beta^\ast(\br)\psi_\alpha(\br)\,,
\end{eqnarray}
$e$ is the electron charge, $\varepsilon_\alpha$ and $\psi_\alpha(\br)$
are the single-particle eigenvalues and eigenfunctions
of the undisturbed system, they are usually calculated
with in a Hartree-Fock or a local-density approximation.
$f(\varepsilon) = \Theta(E_{\rm F}-\varepsilon)$,
and $\gamma>0$ is smaller than $\Delta$.
Within the RPA, the absorption coefficient
(proportional to the energy dissipation per unit time)
may be written as \cite{ber94}
\be
\label{eq:alpha1}
\alpha(\omega) = \frac{\hbar\omega}{2E_0^2} \mbox{Im}\, d(\omega)
\ee
where 
\be
d(\omega) = 
\int\rd\br\rd\brp \delta\varrho^\ast(\br) 
\Pi_0(\br,\br^\prime;\omega)\varphi(\brp)
\ee
is the (complex) dipole moment and 
the asterisk denotes complex conjugation.

In the following we derive an explicit analytical
expression for the absorption
coefficient $\alpha(\omega)$ of a
finite electronic quantum system
in an external electric field, valid
in the frequency range
$\Delta/\hbar <  \omega \ll \omega_{\rm p}$.
We make use of the approximations suggested
in  \cite{wilk00}.
First, according to Fermi's golden rule,
the absorption coefficient is
\be
\label{eq:golden}
\label{eq:alpha2}
\alpha(\omega) \simeq
\frac{\pi\hbar^2 e^2\omega^2}{2\Delta^2 E_0^2}
\Big|\langle \psi_\alpha |\varphi| \psi_\beta\rangle 
\Big|^2_{\mbox{}\hspace*{-2mm}\stackrel{\scriptstyle \varepsilon_\alpha \simeq E_{\rm F}}{\mbox{}\hspace*{3mm}\varepsilon_\alpha\!-\varepsilon_\beta\simeq \omega}}\,.
\ee
Second, the matrix elements of $\varphi$ are evaluated within
a semi-classical approximation
\cite{meh97e,wilk87,eck92,meh95,meh00a}. 
Third, $\varphi$ itself is
determined within a quasi-classical approximation:
according to (\ref{eq:rpa}),
the effective electric potential is given by
(in symbolic notation)
\begin{eqnarray}
\label{eq:varphi}
\varphi &=& -\Pi_0 ^{-1}\delta\varrho\,.
\end{eqnarray}
In order to determine $\delta\varrho$,
eqs. (\ref{eq:rpa}) are usually
solved numerically, using a
 real-space discretisation \cite{ber90} or
 by expanding in a suitable
 basis set. 
An approximate
analytical solution may be obtained
by noting that for $\omega \ll \omega_{\rm p}$,
$||G\Pi_0||\gg 1$. In other words,
$\delta\varrho$  is well approximated 
by the classical charge density $\delta\varrho_{\rm cl}$
of the metallic system subjected to 
an external potential $\varphi_{\rm ext}$,
\be
\label{eq:rhocl}
\label{eq:class}
\Delta\varphi_{\rm cl} = -\delta\varrho_{\rm cl}/\epsilon_0
\ee
with $\varphi_{\rm cl}(\br) \rightarrow \varphi_{\rm ext}(\br)$
as $|\br|\rightarrow\infty$ and $\varphi_{\rm cl} = 0$
within the system (in the classical limit, the external field
is thus screened out completely).
Eq. (\ref{eq:rhocl}) may be solved 
for $\delta\varrho_{\rm cl}$ using standard  methods 
\cite{lan84}.
Fourth, $\Pi_0$ is determined within a quasi-classical
approximation \cite{kir75,wilk00}
\be
\label{eq:quasicl}
\Pi_0(\br,\brp;\omega) = e^2\nu_d \left[ \delta(\br-\brp) + \ri \omega
P^{(d)}(\br,\brp;\omega)\right]
\ee
where  $\nu_d$ is the density of states per unit volume
in $d$ dimensions 
and $P^{(d)}(\br,\brp;\omega)$ is the Fourier transform
of the classical propagator $P^{(d)}(\br,\brp;t)$.
In ballistic systems it is written as a sum over classical 
paths $p$ from $\br$ to $\brp$
\be
P^{(d)}(\br,\brp;\omega) =\!\!\!\!\!\!\! \sum_{{\rm cl. paths}\,p}
\left|\mbox{det}\left[\frac{\partial (\brp)}{\partial (\tau_p,\bbox{n}_p)}
\right]\right|^{-1}\!\!\!\exp(\ri \omega \tau_p)\,.
\label{eq:prop}
\ee
Here $\tau_p$ is the time taken from $\br$ to $\brp$
along the path $p$, and $\bbox{n}_p$ is a unit vector
describing the direction of the initial velocity.
For diffusive systems see e.g. \cite{meh97a}.

In the following we show
by comparison with quantum-mechanical RPA calculations
that (\ref{eq:golden}-\ref{eq:prop}) provide a qualitative
and quantitative description of absorption
in the frequency range $\Delta/\hbar <  \omega \ll \omega_{\rm p}$.

\begin{figure}
\centerline{\psfig{file=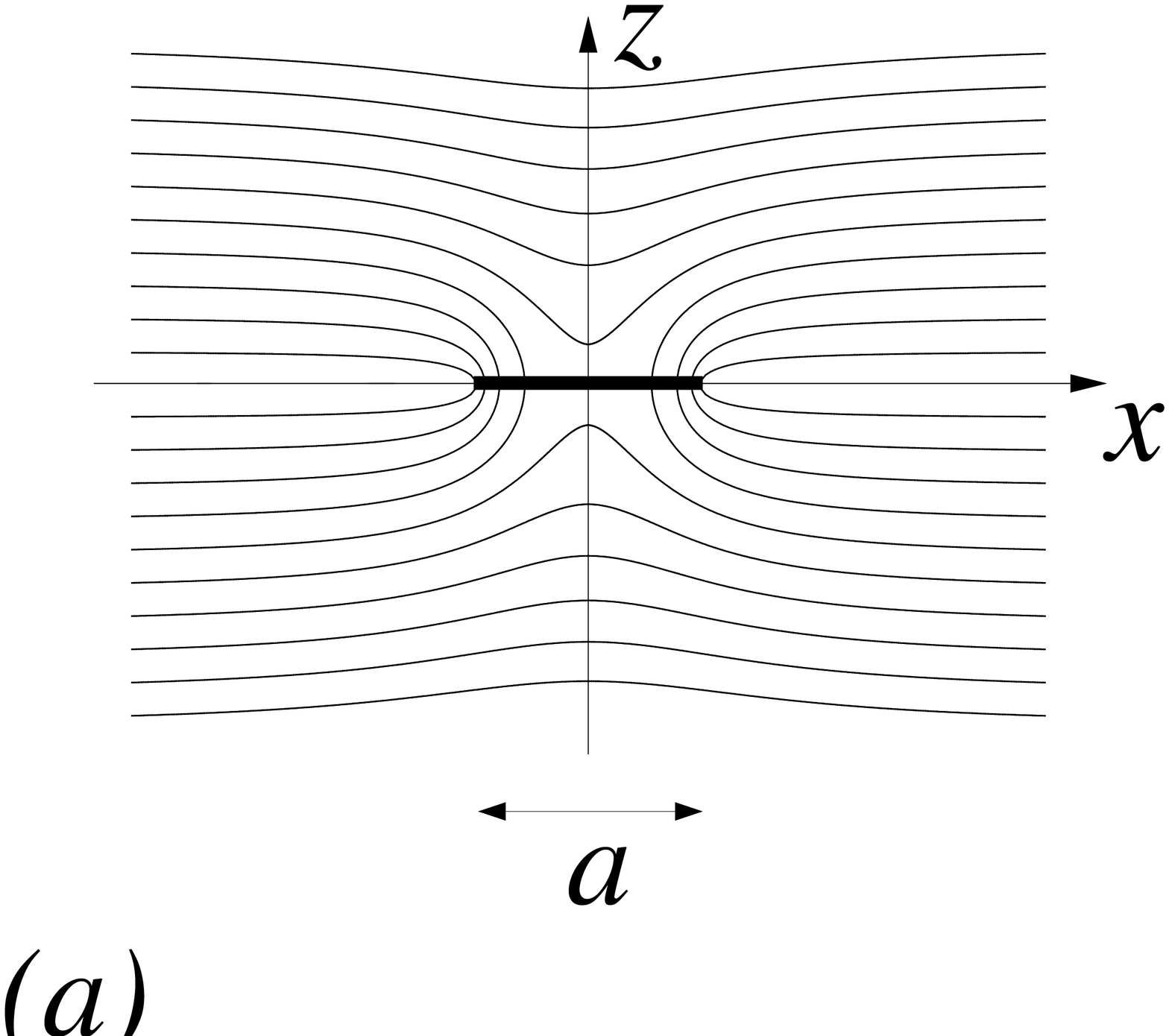,width=3.0cm}\hspace*{0.5cm}
\psfig{file=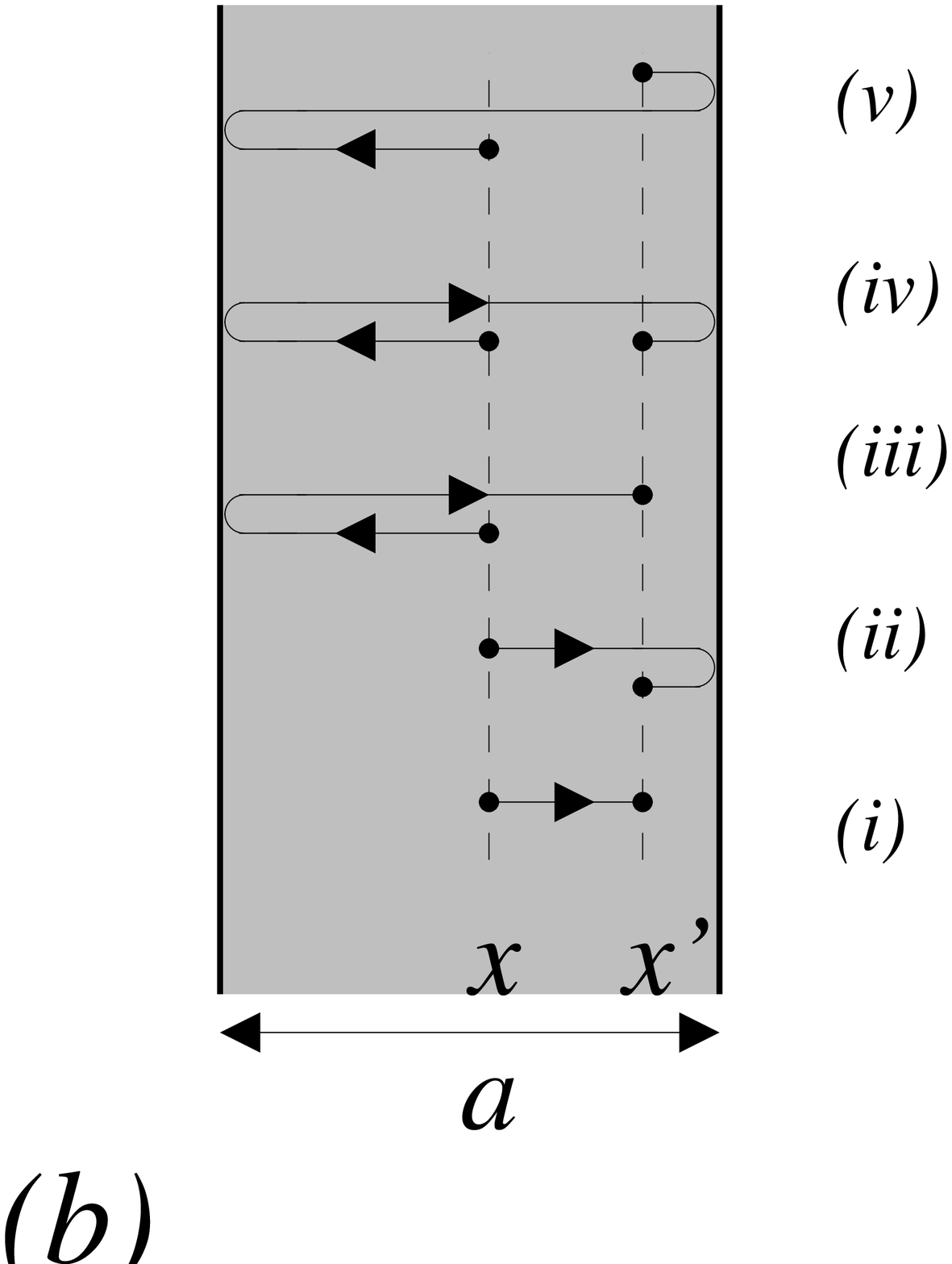,width=2.7cm}}
\vspace*{5mm}
\caption{\label{fig:1} 
(a) Electric field lines (in the $x$-$z$-plane)
for an infinitely long metallic strip of width $a$ 
(in the $x$-$y$-plane, oriented along the $y$-axis)
placed in a constant external electric field
$\bbox{E}_{\rm ext} = -E_0 \hat{\bf e}_x$ .
(b) Classical paths from $x$ to $x^\prime$
contributing to $\Lambda_0(x,x^\prime;\omega)$.}
\end{figure}

{\em Two-dimensional strip.}
We consider a two-dimensional electron gas 
confined to a strip in the $x$-$y$-plane surrounded by vacuum \cite{note},
subject to a time-dependent
electric field $\bbox{E}_{\rm ext}$
directed along the negative $x$-axis [compare fig.~\ref{fig:1}(a)].
The width of the strip (along the $x$-axis) is $a$, its
 length $L$ (along the $y$-axis),
with $L \gg a$.  Within the system, the electrons move ballistically,
and they are specularly reflected at the
boundaries at $x = \pm a/2$.

We write $\delta\varrho(\br) = \delta\sigma(x,y)\,\,\delta(z)$.
For $L \gg a$, the surface-charge density $\delta\sigma$             depends
on $x$ only and the resolvent $G$ is written as
\be
G = 
\frac{1}{\epsilon_0}
\int\!\frac{\rd q}{2\pi}\, \frac{1}{2 |q|}\, {\rm e}^{\ri q (x-x^\prime)}
{\rm e}^{- |q| |z-z^\prime|}\,. 
\ee
With
\be
\frac{1}{L}\int\rd y \rd y^\prime\,\Pi_0(\br;\brp;\omega)
= \Lambda_0(x,x^\prime;\omega)\,\delta(z)\,\delta(z^\prime)\,,
\ee
the RPA-equations (\ref{eq:rpa}) are reduced to a set of 
one-dimensional
equations for $\varphi(x,z\!=\!0)$
with the kernel $\Lambda_0(x,x^\prime;\omega)$.
We model the confinement in the $x$-direction by introducing
hard-wall boundary conditions. This is adequate in the
range of parameters considered below and simplifies
the  quasi-classical analysis.
We solve the resulting self-consistent equations
numerically using a real-space discretisation and obtain
the absorption coefficient from (\ref{eq:alpha1}).

The corresponding quasi-classical approximation
for $\alpha(\omega)$ is obtained
as described above:
the classical surface-charge density
is determined 
by solving (\ref{eq:class}) in elliptic cylinder coordinates:
$\delta\sigma_{\rm cl}(x) = 2\epsilon_0 E_0 (a^2/4-x^2)^{-1/2}
\Theta\left(|x| -a/2\right)$.
The corresponding classical 
field lines are shown in fig.~\ref{fig:1}(a).
According to eqs.~(\ref{eq:quasicl},\ref{eq:prop}) the one-dimensional kernel
$\Lambda_0(x,x^\prime;\omega)$ is given by
a sum over classical paths as shown in fig.~\ref{fig:1}(b)
which may be summed by Poisson summation.
Using (\ref{eq:varphi}) one obtains for
the effective electric potential
\begin{eqnarray}
\label{eq:phi1}
\varphi(x) &=& 
\sum_ {\mu > 0} \varphi_\mu \cos\left[\mu\pi\left(x/a+1/2\right)\right]\,,\\
\varphi_\mu &=&\frac{2\pi\epsilon_0 E_0}{e^2 \nu_2}
\frac{\sqrt{\omega^2\!\!-\!\wc^2\mu^2}}
{\omega\!-\!\!\sqrt{\omega^2\!\!-\!\omega_{\rm c}^2\mu^2}}
\sin\Big(\frac{\mu\pi}{2}\Big)J_1\Big(\frac{\mu\pi}{2}\Big)\,.
\nonumber
\end{eqnarray}
Eq. (\ref{eq:phi1}) has an intuitive interpretation:
neglecting a correction term $\varphi_{\rm bdy}(x)$ which
is small except for $x$ in a boundary layer of width
$\delta x\propto v_{\rm F}/\omega$, eq. (\ref{eq:phi1}) can be written as
a sum over two terms
\be
\label{eq:phi2}
\varphi \simeq \varphi_{\rm stat} + \varphi_{\rm dyn}
\ee
where $\varphi_{\rm stat}(x,z\!=\!0)
= (\epsilon_0\,q_{\rm s})^{-1} \delta\sigma(x)$ 
is the (linearised) TF potential ($q_s = e^2\nu_2/\epsilon_0$
is the two-dimensional TF  screening vector) and
$\varphi_{\rm dyn}$ is a dynamical contribution
corresponding to a current building up the
screening charges. It obeys
$\partial^2 \varphi_{\rm dyn}/\partial x^2
= - (m_e \omega^2/e^2) \,\, \delta\sigma/\sigma_0
$
where $m_e$ is the electron mass and $\sigma_0$ is the areal charge
density of the electrons. 

Fig.~\ref{fig:2}(a) shows $\varphi(x)$
according to (\ref{eq:phi1}) and (\ref{eq:phi2})
compared with the results of a numerical RPA
calculation.  One observes excellent agreement
(and $\varphi_{\rm bdy}$ is small except at the boundary).
Our results  show that for larger frequencies 
$(\omega > \omega_{\rm c})$,
the dynamical potential $\varphi_{\rm dyn}$  makes a significant contribution
to $\varphi$ and dynamical screening effects cannot be neglected.
For the absorption coefficient we obtain using
eqs. (\ref{eq:golden}) and (\ref{eq:phi1}) 
\be
\label{eq:absresult}
\alpha(\omega) \simeq \frac{\pi^2\hbar\epsilon_0^2a}{e^2\nu_2}\frac{\mbox{}\,\omega^2}{\mbox{}\,\wc^2}
\sum_{\mu> [\omega/\wc]}^{\rm odd} \frac{\sqrt{\mu^2\wc^2-\omega^2}}{\mu^2} J_1^2(\mu\pi/2)
\ee
with limiting forms
\be
\label{eq:result}
\alpha(\omega) = 
\left\{
\begin{array}{ll}
 \pi^2 C\hbar\epsilon_0^2 a \omega^2/(e^2 \nu_2 \wc)& \mbox{for $\omega \ll\omega_{\rm c}$}\,,\\
 \pi\hbar \epsilon_0^2 a\omega/(4e^2\nu_2) & \mbox{for $\omega \gg\wc$}
\end{array}
\right .
\ee
and $C\simeq 0.12$. In fig.~~\ref{fig:2}(b), 
we show quantum-mechanical RPA results in comparison
with eqs. (\ref{eq:absresult}) and (\ref{eq:result})
and observe excellent agreement.
We observe
prominent resonances in the absorption coefficient
near odd multiples of $\wc$,
due to single-particle cyclotron orbits
(electrons moving in phase with the external field).
This establishes that the single-particle
resonances conjectured in \cite{aus93} 
exist within the RPA. 
Their positions, strengths and shapes
are very well described by (\ref{eq:absresult}). 
Eq. (\ref{eq:result}) and the inset of fig.~\ref{fig:2}(b) show that
in the limit of large frequencies ($\omega \gg \omega_{\rm c}$),
the absorption coefficient is linear  
in $\omega$. In the opposite limit ($\omega \ll\wc$)
where the TF approach is adequate,
it is quadratic.

 We conclude that the quasi-classical approximation described
 above, for the parameters considered here,
 provides a quantitative description of 
 the optical properties.

\begin{figure}
\centerline{\psfig{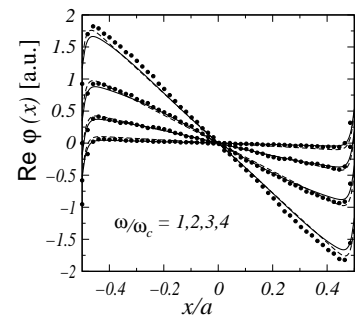}}
\hspace*{0.5cm}
\caption{\label{fig:3} Shows $\varphi(x)$ for
a thin film of width $a=50\au$ with $\rs = 1$:
RPA results ($\bullet$), 
analytical results according
to eq. (\ref{eq:varphi})
(\protect\rule[0.5ex]{0.75cm}{0.15mm})
and using $\varphi\simeq\varphi_{\rm stat} + \varphi_{\rm dyn}$
($- - -$).}
\end{figure}

{\em Three-dimensional thin film.} 
To conclude we discuss $\varphi$ for
a thin film \cite{note2} of width $a$ in the
$y$-$z$-plane subject to an external potential 
$\varphi_{\rm ext}(\br) = E_0 x \hat{\bf e}_x$.
The classical charge density is concentrated
at the boundary, 
$\delta\varrho_{\rm cl}(x) = \pm\delta\sigma\,\delta(x\pm a/2)$.
The corresponding static and dynamical contributions
to $\varphi(x)$ are singular;
we  thus use the TF charge density \cite{new70} instead 
of $\delta\varrho_{\rm cl}$
(appropriate in the limit of small $\ks$ 
corresponding to high electron densities):
$\delta\varrho_{\rm TF}(x) = \ks \epsilon_0 E_0 \sinh(\ks x)/\cosh(\ks a/2)$.
Here  $k_{\rm s}^2 = e^2\nu_3/\epsilon_0$ is the three-dimensional
TF screening vector. 
The RPA equations are easily solved 
within a real-space discretisation.  Our results for $\varphi(x)$ are shown in
fig.~\ref{fig:3}, and compared to results of the analytical approach
using eqs. (\ref{eq:golden}-\ref{eq:prop}).

We have also calculated the absorption coeffcient 
for $\Delta/\hbar < \omega \ll \omega_p$ within the RPA.
The analytical approach must be used with caution
in the case of the film since it requires that
$\varphi$ be smooth on the scale of $\lambda_{\rm F}$.
It turns out that $\alpha(\omega)$ is, to a good approximation, quadratic in $\omega$
as opposed to the two-dimensional case. As in $d=2$ dimensions we observe
resonances near odd multiples of $\wc$ (not shown).  

\narrowtext

\begin{figure}
\centerline{\psfig{file=Fig2a.eps,width=4.2cm,clip=}
	\hspace*{0.1cm}
\psfig{file=Fig2b.eps,width=3.9cm,clip=}}
\vspace*{0.2cm}
\caption{\label{fig:2} 
Left: $\varphi(x)$ for
a strip of width $a=10^3 \au$ with $\rs = 1$:
quantum-mechanical results ($\bullet$),
analytical results according
to eq. (\ref{eq:phi1}) 
(\protect\rule[0.5ex]{0.75cm}{0.15mm})
and eq. (\ref{eq:phi2}) 
($- - -$). The inset shows the correction term $\varphi_{\rm bdy}(x)$.
 Right: Shows $\protect\mbox{Im}\, d(\omega)$ 
 (for $a=10^4 \au$ and $\rs = 1$) as a function
 of $\omega$:  RPA result
(\protect\rule[0.5ex]{0.75cm}{0.15mm})
 and eq.~(\protect\ref{eq:absresult}) ($- - -$).
 The inset shows $\alpha(\omega)$ as a function
 of $\omega$: RPA result (\protect\rule[0.5ex]{0.75cm}{0.15mm})
and eq.~(\protect\ref{eq:result}) ($- - -$). 
As usual
$\rs=r_0/a_0$ where $a_0$ is the Bohr radius and
$r_0$ is the length scale defined in terms of
area per electron. }
\end{figure}

\end{multicols}

\begin{thebibliography}{xxx}
\bibitem{note} The influence of dielectrics above and below
the strip, or of grounded electrodes are easily
incorporated, see for instance \cite{fet86}.
\bibitem{note2} Collective effects and the nature of the plasma resonances 
in thin films were discussed in \cite{fer58} and, within
the RPA, in \cite{kan61,new70}.
\end{thebibliography}
\end{document}